WILEY-VCH

**Dipolar Switching of Charge-injection Barriers at Electrode/Semiconductor Interfaces as a Mechanism for Water-induced Instabilities of Organic Devices**

*Ryo Nouchi\**

R. Nouchi, Prof.
Department of Physics and Electronics, and Nanoscience and Nanotechnology Research Center, Osaka Prefecture University, Sakai 599-8570, Japan
PRESTO, Japan Science and Technology Agency, Kawaguchi 332-0012, Japan
E-mail: r-nouchi@pe.osakafu-u.ac.jp

An electrode-contact-related mechanism for the operational instability of organic electronic devices is proposed and confirmed via observation of a water-induced change in charge-injection barrier heights at the electrode/organic semiconductor interfaces. Water molecules in air penetrate into the organic crystal via diffusion, and an external electric field orients the electric dipole of the water molecules at the electrode surfaces, leading to dipolar switching of the charge-injection barrier height. As a result of the switching, current–voltage curves of two-terminal Au–rubrene–Au devices change from symmetric to asymmetric, showing diode-like rectification and reversible switching of the diode polarity. The device shows the highest current switching ratio of 267 for the switching voltage of 3 V, corresponding to an electrode work function change of > 144 meV. The mechanism proposed herein will be important especially for short-channel organic devices, which are indispensable for applications such as organic integrated circuits.

**1. Introduction**

Realistic application of electronic devices requires stable operation for a prolonged period of time in the environment in which the devices are used. In many cases, the devices are used in ambient air. Thus, the devices are always exposed to atmospheric molecules such as nitrogen, oxygen and water. Among them, water is known to largely affect the operation of electronic devices, and various mechanisms have been proposed for dealing with water-induced instabilities.[1-5]

In particular, electronic devices based on organic or molecular semiconductors are generally weak against water-induced instabilities. This is due to a weak intermolecular interaction or large intermolecular separation, which allows small molecules such as water to easily penetrate into the crystal. Considering a field-effect transistor configuration, the region around the channel, where the charge conduction occurs, would be largely affected by the water penetration. The proposed mechanisms include diffusion of water-related species into the gate dielectric,[6-8] water-induced polarization of the gate dielectric,[9, 10] and charge carrier trapping at/near the channel.[11-15] All these mechanisms are related to the gate dielectric itself or the interface with the gate dielectric. However, water molecules are expected to also reach the source/drain electrode contacts. Nevertheless, an electrode-contact-related mechanism of water-induced instability in organic devices has not been reported so far.

In this paper, we propose a mechanism involving the electrode contacts and water molecules, and experimentally confirm it via a change in the current–voltage ($I$–$V$) characteristics of a short-channel two-terminal (electrode–organic-semiconductor–electrode) device. The change is manifested as a reversible switching of the polarity of diode-like $I$–$V$ characteristics, which is attributable to a change in the charge-injection barrier height at the source/drain electrode contacts. The switching behavior is not observed in pure nitrogen or pure oxygen environments but is observed in a water-containing environment. The electrode-contact-related phenomena are generally more dominant in shorter inter-electrode spacings, i.e., shorter channel lengths. Therefore, the mechanism proposed and confirmed herein will be significant for application in microscale devices, which are important for constructing organic integrated circuits[16-18] and related architectures that require miniaturization.

## 2. Results

**Figure 1**a depicts the mechanism of water-induced instability proposed herein. A weak intermolecular interaction in molecular solids results in a relatively large intermolecular spacing. Thus, foreign small molecules such as water can easily penetrate into the molecular crystal. In addition, if the molecular layer is a polycrystalline film, the penetration is expected to be largely enhanced owing to the presence of grain boundaries. When the penetrated water molecules diffuse until reaching the source/drain electrode contacts, the water molecules are expected to align their electric permanent dipole to the electric field generated by the drain voltage. Subsequently, the aligned water molecules form an electric double layer at the electrode contacts as shown in Figure 1b, which is expected to lead to the modulation of the effective work function of the source/drain electrodes.[19, 20] The direction of the expected change in the work function is opposite in the source and drain contacts. Thus, as a result of the dipolar switching, asymmetry in the charge-injection barrier height is introduced to the device, which can be regarded as a two-terminal metal–semiconductor–metal diode with asymmetric Schottky barrier heights.[21]

Furthermore, the resultant asymmetry in the charge-injection barrier heights can be reversed by changing the polarity of the drain voltage. The opposite direction of the electric field arising from the polarity-reversed drain voltage leads to oppositely aligned dipoles of water molecules at the electrode contacts as shown in Figure 1c. This leads to opposite asymmetry in the charge-injection barrier heights. Thus, the polarity of the diode-like rectification behavior in $I$–$V$ characteristics is also reversed as was observed in a metal–organic semiconductor–metal diode with a switchable molecular modification layer on the electrode surface.[22-24]

**Figure 2**a shows a schematic diagram of a measured device consisting of a rubrene single crystal bridged over two 14-nm-thick Au electrodes with a very short (< 1.0 μm)

spacing, where the doped Si substrate and the left (source) terminal were grounded, and the (drain) voltage $V$ was applied for inducing the switching and also for characterizing the diode. The measurement scheme for the characterization of the reversible switching behavior was as follows: (1) initial characterization of an $I–V$ curve within $V = \pm1$ V; (2) induction of switching by applying higher positive voltage of $V = +1, +3, +5, +7,$ or $+9$ V for 60 s; (3) characterization again within $V = \pm1$ V after application of the positive switching voltage; (4) induction of switching by applying negative higher voltage of $V = −1, −3, −5, −7,$ or $−9$ V for 60 s; (5) characterization again within $V = \pm1$ V after application of the negative switching voltage; (6) repeat steps (2) to (5) four times (five times in total for each switching voltage).

Figure 2b illustrates the reversible switching behavior measured in a humidity-controlled environment (32.5%). The initial $I–V$ curve is naturally symmetric because no asymmetry was intentionally introduced in the device fabrication procedure. However, after applying switching voltage of $+3$ V, the symmetric $I–V$ curve was changed to an asymmetric one where apparent rectification behavior was observed. Subsequently, switching voltage of opposite polarity ($−3$ V) was applied, after which the polarity of the rectifying $I–V$ curve was reversed. Only the first three steps of the series of switching measurements are shown in Figure 2b, but the switching behavior was further confirmed via the consecutive application of positive and negative switching voltages as shown in Figure 2c.

The reversible switching behavior in $I–V$ characteristics is understood through the energy diagrams displayed in Figure 2d. First, by considering the values of the work function of polycrystalline Au films (5.1 eV)[25] and the valence band maximum of rubrene (4.85 eV),[26] it is concluded that holes are the majority carriers of the two-terminal device fabricated in this study. Thus, only the hole-injection barrier is considered in the energy diagrams. In the initial $I–V$ curve, the current level is moderate and symmetric, and thus, moderate hole-injection barriers exist at both electrode contacts. After the application of the positive switching voltage, the current level is higher in the positive $V$ region than in the negative region, which indicates that the hole-injection barrier is lower in the right (drain) contact than in the left (source) contact. After the application of the negative switching voltage, the current asymmetry was reversed, and the asymmetry in the barrier height was also reversed. As shown in these energy diagrams, the reversible switching of the polarity of the diode-like rectification behavior is understood through the reversible switching of the asymmetry in the hole-injection barrier heights at the two electrode contacts.

As a measure of the extent to which the current changes, the current switching ratio is defined as follows. As shown in Figure 2b, the absolute current increased in the positive (negative) $V$ region after the application of the positive (negative) switching voltage. Thus, the ratio between the absolute currents measured at the positive (negative) $V$ region after and before the application of the positive (negative) switching voltage indicates how much the absolute current increased after the switching. The ratio of the absolute current measured at $V$ = +1 or −1 V was used for calculating the switching ratio. In the series of switching measurements, the number of applications of switching voltages for each absolute switching voltage was ten in total (five for positive and five for negative switching voltages). The average switching ratio of the ten switching procedures was obtained.

**Figure 3**a illustrates the measurement of the current switching ratio in humidity-controlled air. The measurements were performed with a device with an electrode spacing of 0.6 μm under relative humidity from 25% to 62.5% in steps of 7.5%. At each humidity, the switching voltage was applied in the order $V$ = ±1, ±3, ±5, ±7, and ±9 V. Even at the lowest humidity tested (25%), the switching ratio first increased as the switching voltage increased from 1 V to 5 V, and thereafter slightly decreased as the switching voltage increased beyond 5 V. This can be understood through the concurrent presence of the gate-related instability, especially owing to the generally observed current reduction via charge carrier trapping

at/near the channel.[11-15] In the present study, the gate terminal as well as the source terminal were grounded, but the application of the drain voltage generates a finite gate electric field near the drain end. The finite gate field reduced the current through the gate-related instability. At each switching event, the switching ratio is defined as the ratio of the absolute current after to before the switching. Because the gate-related instability during the switching voltage application reduced the absolute current after the switching, the switching ratio decreased as a result of the gate-related instability. Apart from the influence of the gate-related mechanism, the switching ratio tended to increase as the relative humidity increased from 25% to 47.5%, which can be attributed to the dipolar switching of charge-injection barrier heights.

Figure 3a also shows that the switching ratio started to decrease as the relative humidity increased beyond 47.5%, and almost no switching was observed at 62.5%. This can be attributed to the irreversible degradation of the rubrene crystal, which is possibly due to electrochemical oxidation. The oxidized channel became almost insulating as shown in Figure 3b, and thus, switching behaviors could not be observed. The irreversibility was further confirmed through a series of measurements with a different sequence. Figure 3c shows the evolution of absolute current measured at $V = +1$ or $-1$ V after the application of each switching voltage. The inter-electrode spacing of the device used for the measurements was 0.8 μm. The sequence of setting of the relative humidity is shown in the explanatory notes of each plot, namely, in the order (1) 27.5%, (2) 50%, (3) 27.5%, (4) 72.5%, and (5) 27.5%. During the first three sequences, the change in the current level was not large. However, during the switching measurements at the highest humidity tested (72.5%), the current level rapidly decreased to almost zero, and the current level did not recover after setting back to the low relative humidity of 27.5%. These results confirm that the current decrease through the application of switching voltage in high humidity conditions was caused by irreversible degradation of the rubrene crystal.

The necessity of $H_2O$ for switching was further confirmed through measurements in a gas-controlled environment as shown in **Figure 4**a. The switching ratio was determined using a method similar to that used in Figure 3. Figure 4a illustrates the measurement of the current switching ratio in a gas-controlled environment: $N_2$ including 400 ppm of $H_2O$ (filled circle), $N_2$ (open circle), and $O_2$ (triangle). The channel length of the measured device was 0.4 μm. The switching ratios are very low for $N_2$ and $O_2$ environments, indicating that the charge-injection barrier height was subject to almost no switching in these environments. In contrast, inclusion of a small amount of $H_2O$ molecules drastically enhanced the switching, where the switching ratio was determined to be as high as ca. 30 for the switching voltage of 9 V. These results demonstrated the necessity of $H_2O$ for the switching of the charge-injection barrier.

In the $N_2$ environment with a small amount of $H_2O$, the switching ratio shows a monotonic increase as the switching voltage increases below 9 V. This behavior is different from the nonmonotonic behavior in Figure 3 and is considered to arise from the difference in the concentration of $H_2O$ molecules. Actually, 400 ppm of $H_2O$ (Figure 4) corresponds to the absolute humidity of 0.36 g m$^{-3}$, which is much lower than that (6.4 g m$^{-3}$) corresponding to the relative humidity of 25% at 27 °C. In the investigated system with a very short channel length, the contact-related mechanism proposed herein is dominant in the very-low-humidity range, but the gate-related mechanism shows a large contribution under the humidity range realistic in our place of living.

It should be noted that the channel lengths of the measured devices in Figures 2 to 4 are not identical, which is due to a technical reason (a failure in a lithographic process to prepare devices with the identical channel length). The different channel length should change the profile of an electric field strength along the channel. However, the normally on nature of the rubrene devices (the finite current without application of the gate voltage) indicates that the channel resistance is not so large, and instead the electrode contact resistance dominates the overall device resistance. Thus, the voltage drop occurs mostly at the electrode/organic-semiconductor interfaces where the dipole of water molecules switches. Therefore, the dipolar switching is not so much affected by the difference in the channel length.

## 3. Discussions

As shown above, the polarity reversal of the rectifying *I–V* curve is attributable to the water-induced reversible switching of the charge-injection barrier height, but a microscopic view that can explain the experimentally obtained energy diagrams in Figure 2d is not provided. First, the electric field generated by the application of the (drain) voltage aligns the dipole of $H_2O$ as shown in Figure 5a. The direction of the aligned dipole is opposite at the left (source) and right (drain) contacts, leading to the asymmetry in these two contacts. Subsequently, we consider how the aligned $H_2O$ molecules change the work functions of the electrodes. Intuitively, the direction of the switching could be determined by an electric double layer formed by the permanent electric dipole[19, 20] of $H_2O$ molecules, from which energy diagrams were deduced as shown in Figure 5b. It is easily understood that the diagrams are different from the experimentally obtained ones in Figure 2d.

Thus, a microscopic view that does not rely on the dipole model is necessary to explain the observed direction of the switching. Another possible feature of $H_2O$ adsorption is the so-called push-back (or "pillow") effect.[27-29] At the bare metal surface, the electronic cloud spills out from the metal into the vacuum, which forms an electric double layer with the negative pole facing outward.[30] When foreign atoms or molecules ($H_2O$ in the present case) are adsorbed on the electrode metal surface, the metal electronic cloud in the vacuum side is partly pushed back into the metal via the Pauli repulsion from the electronic cloud of the adsorbate. This leads to a reduction in the electric double layer at the metal surface, resulting in a decrease in the work function. By considering the extent of the electronic cloud of $H_2O$ molecules,[31] the decrease in the work function is considered to be larger when the hydrogen side faces the electrode metal surface, from which energy diagrams were deduced as shown in Figure 5c. These diagrams are consistent with the experimentally deduced ones in Figure 2d, and thus, the push-back model properly explains the results. Although quantitative examinations are absent at present, this finding qualitatively indicates that the change in the effective work function caused by the push-back effect was larger than that caused by the dipole effect.

Two-terminal metal–organic-semiconductor–metal devices with short inter-electrode spacings can be treated as two Schottky diodes connected back-to-back in series.[21, 32-36] The $I–V$ characteristics of such devices show a rapid increase in $|I|$ within a small $|V|$ region and subsequent saturation of $|I|$ in a higher $|V|$ region, which was indeed observed as shown in the top panel of Figure 4b. Even if asymmetry is introduced into the charge-injection barriers at the two electrode contacts, the $I–V$ characteristics have been predicted to follow the same trend, but the current level after saturation in the positive and negative $V$ regions becomes different depending on the degree of asymmetry in the charge-injection barrier height.[21] In the center and bottom panels of Figure 4b, which were measured after applying switching voltages, the $I–V$ curves show a quasi-saturation behavior, and the current levels are different in the positively and negatively biased regions. Thus, the results shown in Figure 4 can be considered to be within the theoretical framework of two Schottky diodes, with asymmetric Schottky barrier heights, that are connected back-to-back in series.[21]

The $I–V$ characteristics of such systems can be understood through the thermionic emission of charge carriers from an electrode into a semiconductor. The thermionic emission current of a single Schottky diode with a Schottky barrier height of $\Phi_B$ is written as follows:[37]

$$I = AA^*T^2 \exp\left(-\frac{q\Phi_B}{kT}\right)\left[\exp\left(\frac{q(V-IR_S)}{nkT}\right) - 1\right]$$

$$\approx \begin{cases} AA^*T^2 \exp\left(-\frac{q\Phi_B}{kT}\right)\exp\left(\frac{q(V-IR_S)}{nkT}\right) \text{ (for highly positive } V) \\ -AA^*T^2 \exp\left(-\frac{q\Phi_B}{kT}\right) \text{ (for highly negative } V) \end{cases} \tag{1}$$

where $A$ is the cross-section of the current flow path, $A^*$ is the effective Richardson constant, $T$ is the absolute temperature, $q$ is the electronic charge, $k$ is the Boltzmann constant, $R_s$ is the series resistance mainly arising from the semiconductor's bulk resistance, and $n$ is the ideality factor of the thermionic emission behavior. The expression for $I$ for highly negative $V$ is called the reverse saturation current. In the case of two Schottky diodes connected back-to-back in series, the energy barriers at both electrode contacts should be considered. The $I$-$V$ characteristics are now different from Equation (1), and $I$ in both positively and negatively biased regions is governed by the reverse saturation current.[21] If the charge-injection barrier heights are different in the two electrode contacts, $|I|$ in the $V$ region with one polarity (positive or negative) is higher than that in the region with the other polarity, which are determined by the reverse saturation current of a single barrier diode with lower and higher $\Phi_B$, respectively.[21] The application of a switching voltage reversibly switches the $\Phi_B$ values as the interface with higher (lower) $\Phi_B$ values becomes that with lower (higher) $\Phi_B$ values. Therefore, the switching ratio $R_{sw}$ is determined by the difference between the higher and lower heights, $\delta\Phi_B \equiv \Phi_B^{high} - \Phi_B^{low}$, as follows:[24]

$$R_{sw} = \frac{AA^*T^2 \exp\left(-\frac{q\Phi_B^{low}}{kT}\right)}{AA^*T^2 \exp\left(-\frac{q\Phi_B^{high}}{kT}\right)} = \exp\left(\frac{q\delta\Phi_B}{kT}\right). \tag{2}$$

Therefore, the switching ratio obtained from the $I$–$V$ curves that show a quasi-saturation behavior as in Figure 4b can be related to the work function change through the switching, $\delta\Phi_B$, using the relationship in Equation (2). The inset in Figure 4a shows the $\delta\Phi_B$ values calculated using Equation (2) from the experimentally obtained switching ratios for the $H_2O$-containing $N_2$ environment. The work function change of ca. 92 meV was obtained for the switching voltage of 9 V. However, the saturation behavior of the $I$–$V$ curves in Figure 2b is too weak to rationalize the relationship in Equation (2), which is possibly due to a rather large series resistance[38] (mainly, the bulk resistance of the semiconductor layer). Therefore, the $\delta\Phi_B$ values calculated from Equation (2) are expected to be the lower limit of the actual work function change. In the case of the highest switching ratio obtained in this study (ca. 267 for the switching voltages of 3 and 5 V under the relative humidity of 47.5%), the work function change can be estimated to be higher than 144 meV.

### 4. Conclusion

A mechanism of water-induced instability in organic electronic devices was proposed. This mechanism is related to the source/drain electrode contacts and is different from the gate-dielectric-related mechanisms reported so far. The mechanism was confirmed via a change in the current–voltage ($I$–$V$) characteristics of a short-channel two-terminal (electrode–organic-semiconductor–electrode) device. The change was manifested as a reversible switching of the polarity of diode-like $I$–$V$ characteristics, which is attributable to a change in the charge-injection barrier height at the source/drain electrode contacts. The switching behavior was not observed in pure nitrogen or pure oxygen environments but was observed in a water-containing environment, which provides cogent evidence for the proposed mechanism.

This kind of contact-related mechanism is expected to be important in short-channel devices where the contribution of electrode contacts is inherently large. Therefore, the instability mechanism proposed and confirmed herein will be essential, especially in organic integrated circuits and related architectures that require miniaturization. Moreover, the contact-related mechanism is not restricted to organic semiconductors. It is expected to work generally in loosely packed semiconductors, which include nanowire/nanosheet network films in addition to van der Waals molecular solids.

## 5. Experimental Section

*Sample preparation*: A highly doped Si wafer with a thermally oxidized layer was used as a substrate. The thickness of the $SiO_2$ layer was 300 nm. Conventional electron beam lithography processes were used to fabricate electrode patterns on the substrate. The electrodes were 1.0-µm-wide and were made of 14-nm-thick gold with a thin (1-nm-thick) Cr under-layer, which acted as an adhesion layer to $SiO_2$. The inter-electrode spacings were as short as less than 1 µm, which ensures the large contribution of the electrode contacts to the overall electrical resistance of the fabricated devices. Finally, a rubrene single crystal synthesized via physical vapor transport[39, 40] was manually laminated onto the fabricated electrode pattern.

*Measurements*: The electrical characteristics of the fabricated devices were measured using a semiconductor device analyzer (Keysight, B1500A) at room temperature (~27 °C) in a dark condition. Two types of controls of the measurement environment were tested: humidity-controlled air (relative humidity from 25% to 72.5%) and gas-controlled environment ($N_2$, $O_2$, or $N_2$ + 400-ppm $H_2O$). These measurement environments were maintained at ambient pressure (~$10^5$ Pa) in both cases.

## Acknowledgements

This work was supported in part by the Special Coordination Funds for Promoting Science and Technology from the Ministry of Education, Culture, Sports, Science and Technology of Japan; and the Inamori Foundation.

Received: ((will be filled in by the editorial staff))
Revised: ((will be filled in by the editorial staff))
Published online: ((will be filled in by the editorial staff))

References

[1]     M. Egginger, S. Bauer, R. Schwödiauer, H. Neugebauer, N. S. Sariciftci, *Monatsh. Chem.* **2009**, *140*, 735.

[2]     H. Sirringhaus, *Adv. Mater.* **2009**, *21*, 3859.

[3]     S. H. Jin, A. E. Islam, T.-I. Kim, J.-H. Kim, M. A. Alam, J. A. Rogers, *Adv. Funct. Mater.* **2012**, *22*, 2276.

[4]     D. K. Kim, Y. Lai, T. R. Vemulkar, C. R. Kagan, *ACS Nano* **2011**, *5*, 10074.

[5]     M. Lafkioti, B. Krauss, T. Lohmann, U. Zschieschang, H. Klauk, K. V. Klitzing, J. H. Smet, *Nano. Lett.* **2010**, *10*, 1149.

[6]     N. D. Young, A. Gill, *Semicond. Sci.  Technol.* **1992**, *7*, 1103.

[7]     S. J. Zilker, C. Detcheverry, E. Cantatore, D. M. de Leeuw, *Appl. Phys. Lett.* **2001**, *79*, 1124.

[8]     Y. H. Noh, S. Young Park, S.-M. Seo, H. H. Lee, *Organ. Electron.* **2006**, *7*, 271.

[9]     T. Jung, A. Dodabalapur, R. Wenz, S. Mohapatra, *Appl. Phys. Lett.* **2005**, *87*, 182109.

[10]    N. V. V. Subbarao, M. Gedda, P. K. Iyer, D. K. Goswami, *Org. Electron.* **2016**, *32*, 169.

[11]    W. Kim, A. Javey, O. Vermesh, Q. Wang, Y. Li, H. Dai, *Nano. Lett.* **2003**, *3*, 193.

[12]    H. L. Gomes, P. Stallinga, M. Cölle, D. M. de Leeuw, F. Biscarini, *Appl. Phys. Lett.* **2006**, *88*, 082101.

[13]    S. G. J. Mathijssen, M. Cölle, H. Gomes, E. C. P. Smits, B. de Boer, I. McCulloch, P. A. Bobbert, D. M. de Leeuw, *Adv. Mater.* **2007**, *19*, 2785.

[14]    G. Gu, M. G. Kane, *Appl. Phys. Lett.* **2008**, *92*, 053305.

[15]    S. H. Kim, H. Yang, S. Y. Yang, K. Hong, D. Choi, C. Yang, D. S. Chung, C. E. Park, *Org. Electron.* **2008**, *9*, 673.

[16]    G. H. Gelinck, H. E. Huitema, E. van Veenendaal, E. Cantatore, L. Schrijnemakers, J. B. van der Putten, T. C. Geuns, M. Beenhakkers, J. B. Giesbers, B. H. Huisman, E. J. Meijer,

E. M. Benito, F. J. Touwslager, A. W. Marsman, B. J. van Rens, D. M. de Leeuw, *Nat. Mater.* **2004**, *3*, 106.

[17]    E. Cantatore, T. C. T. Geuns, G. H. Gelinck, E. van Veenendaal, A. F. A. Gruijthuijsen, L. Schrijnemakers, S. Drews, D. M. de Leeuw, *IEEE J. Solid-State Circuits* **2007**, *42*, 84.

[18]    A. Yamamura, H. Matsui, M. Uno, N. Isahaya, Y. Tanaka, M. Kudo, M. Ito, C. Mitsui, T. Okamoto, J. Takeya, *Adv. Electron. Mater.* **2017**, *3*, 1600456.

[19]    I. G. Hill, A. Rajagopal, A. Kahn, *J. Appl. Phys.* **1998**, *84*, 3236.

[20]    X. Crispin, V. Geskin, A. Crispin, J. Cornil, R. Lazzaroni, W. R. Salaneck, J.-L. Brédas, *J. Am. Chem. Soc.* **2002**, *124*, 8131.

[21]    R. Nouchi, *J. Appl. Phys.* **2014**, *116*, 184505.

[22]    R. Nouchi, Y. Kubozono, *Org. Electron.* **2010**, *11*, 1025.

[23]    R. Nouchi, M. Shigeno, N. Yamada, T. Nishino, K. Tanigaki, M. Yamaguchi, *Appl. Phys. Lett.* **2014**, *104*, 013308.

[24]    R. Nouchi, T. Tanimoto, *ACS Nano* **2015**, *9*, 7429.

[25]    H. B. Michaelson, *J. Appl. Phys.* **1977**, *48*, 4729.

[26]    Y. Nakayama, S. Machida, T. Minari, K. Tsukagishi, Y. Noguchi, H. Ishii, *Appl. Phys. Lett.* **2008**, *93*, 173305.

[27]    N. D. Lang, *Phys. Rev. Lett.* **1981**, *46*, 842.

[28]    P. S. Bagus, V. Staemmler, C. Woll, *Phys. Rev. Lett.* **2002**, *89*, 096104.

[29]    G. Witte, S. Lukas, P. S. Bagus, C. Wöll, *Appl. Phys. Lett.* **2005**, *87*, 263502.

[30]    N. D. Lang, W. Kohn, *Phys. Rev. B* **1970**, *1*, 4555.

[31]    R. F. W. Bader, G. A. Jones, *Can. J. Chem.* **1963**, *41*, 586.

[32]    X.-L. Tang, H.-W. Zhang, H. Su, Z.-Y. Zhong, *Physica E* **2006**, *31*, 103.

[33]    T. Nagano, M. Tsutsui, R. Nouchi, N. Kawasaki, Y. Ohta, Y. Kubozono, N. Takahashi, A. Fujiwara, *J. Phys. Chem. C* **2007**, *111*, 7211.

[34]    D. R. White, M. Arai, A. Bittar, K. Yamazawa, *Int. J. Thermophys.* **2007**, *28*, 1843.

[35]    A. S. Molinari, I. Gutiérrez Lezama, P. Parisse, T. Takenobu, Y. Iwasa, A. F. Morpurgo, *Appl. Phys. Lett.* **2008**, *92*, 133303.

[36]    A. J. Chiquito, C. A. Amorim, O. M. Berengue, L. S. Araujo, E. P. Bernardo, E. R. Leite, *J. Phys. Condens. Matter.* **2012**, *24*, 225303.

[37]    S. M. Sze, K. K. Ng, *Physics of Semiconductor Devices, 3rd ed.*, John Wiley & Sons, Inc., Hoboken, NJ, USA 2006.

[38]    J. Osvald, *phys. status solidi (a)* **2015**, *212*, 2754.

[39]    R. W. I. de Boer, M. E. Gershenson, A. F. Morpurgo, V. Podzorov, *phys. status solidi (a)* **2004**, *201*, 1302.

[40]    C. Reese, Z. Bao, *Mater. Today* **2007**, *10*, 20.

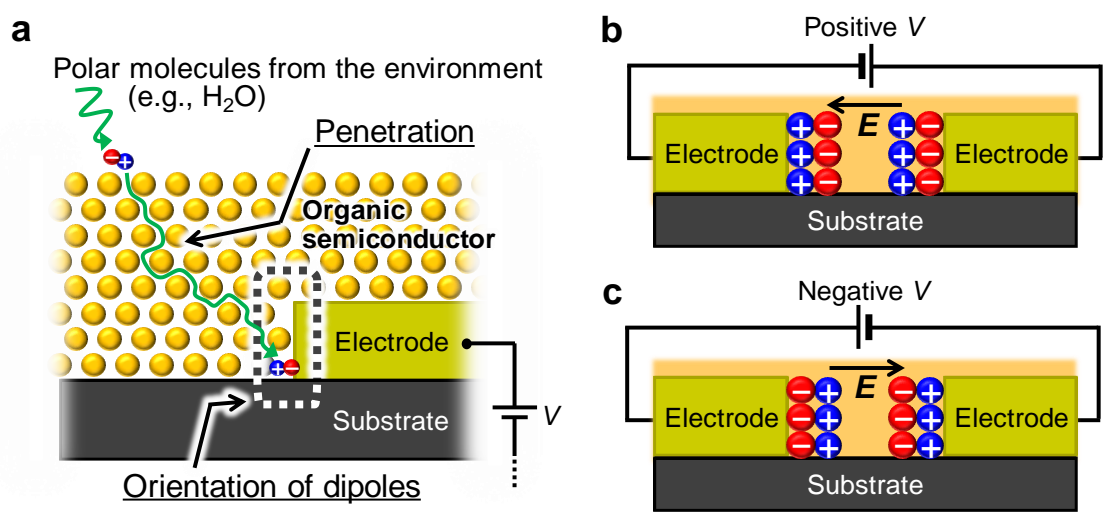

**Figure 1.** Proposed electrode-contact-related mechanism of water-induced instability of organic electronic devices. a) Instability process caused by water penetration into an organic crystal and water dipole orientation at electrode contact surfaces. Directions of water dipoles oriented by the application of b) positive and c) negative voltages.

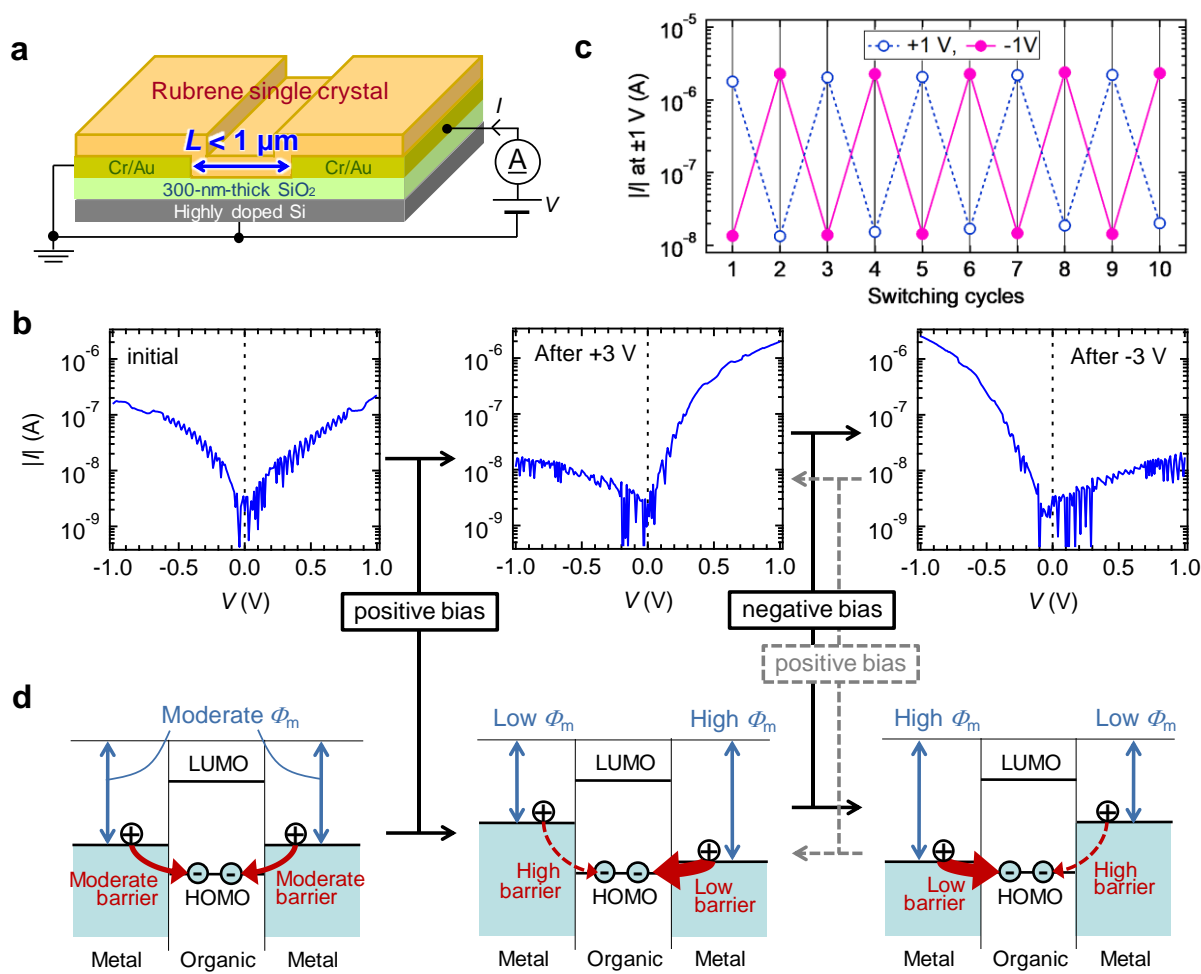

**Figure 2.** Switching properties of *I–V* characteristics of two-terminal organic devices with symmetric electrodes. a) Schematic diagram of Au–rubrene–Au devices fabricated in this study. b) Change of the *I–V* curve from a symmetric one (left) to a rectified one after applying the switching voltage of +3 V for 60 s (center), followed by polarity reversal of the diode-like curve by applying the switching voltage with an opposite polarity (−3 V) for 60 s (right). The channel length of the measured device was 0.6 μm, and the measurements were performed at 27 °C with the relative humidity of 32.5%. c) Absolute current measured at ±1 V after each application cycle of the switching voltage (+3, −3, …, −3 V). The reversible nature of the switching can be observed. d) Energy diagrams constructed from the results in (b). $\Phi_m$ represents the effective work function of the metal electrodes.

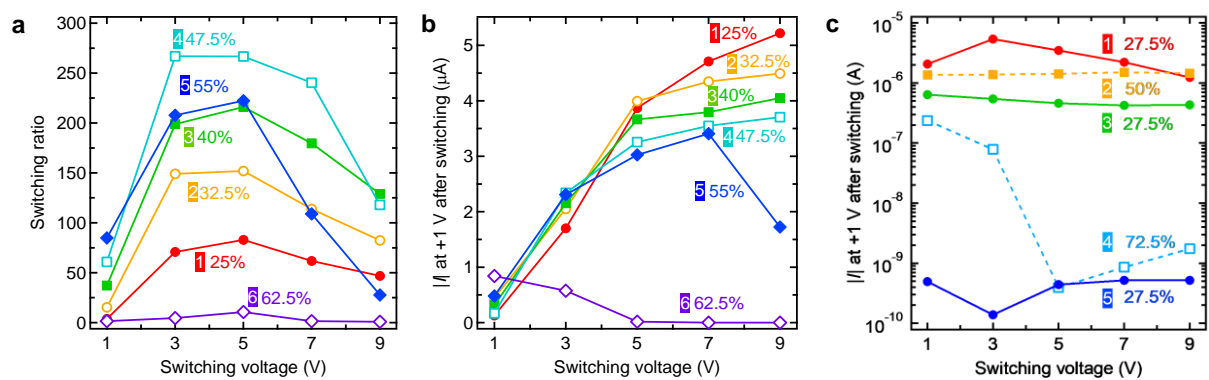

**Figure 3.** Humidity dependence of the switching properties. The numbers shown with a white letter indicate the sequence of measurements. Switching-voltage dependence of a) switching ratios and b) absolute currents at +1 V measured under different values of relative humidity. The device used for the measurements in a) and b) was identical to that shown in Figures 2b and 2c, and its channel length was 0.6 μm. c) Effect of the measurement sequence on the current level. The device used for this measurement had the channel length of 0.8 μm. The current level decreased to the noise floor after measurements at 72.5%.

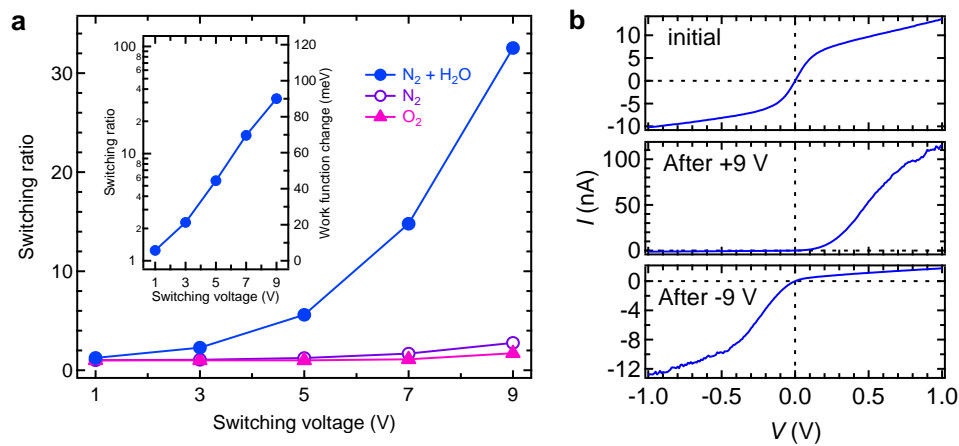

**Figure 4.** Gas dependence of the switching properties. a) Switching-voltage dependence of switching ratios under different gas environments. The corresponding change in the work function is shown in the inset for the H₂O-containing environment. b) Example of *I–V* characteristics measured in the H₂O-containing environment. The *I–V* curve changed from a symmetric one (top) to a rectified one after applying the switching voltage of +9 V for 60 s (center), followed by polarity reversal of the diode-like curve by applying the switching

voltage with an opposite polarity (−9 V) for 60 s (bottom). The channel length of the measured device was 0.4 µm.

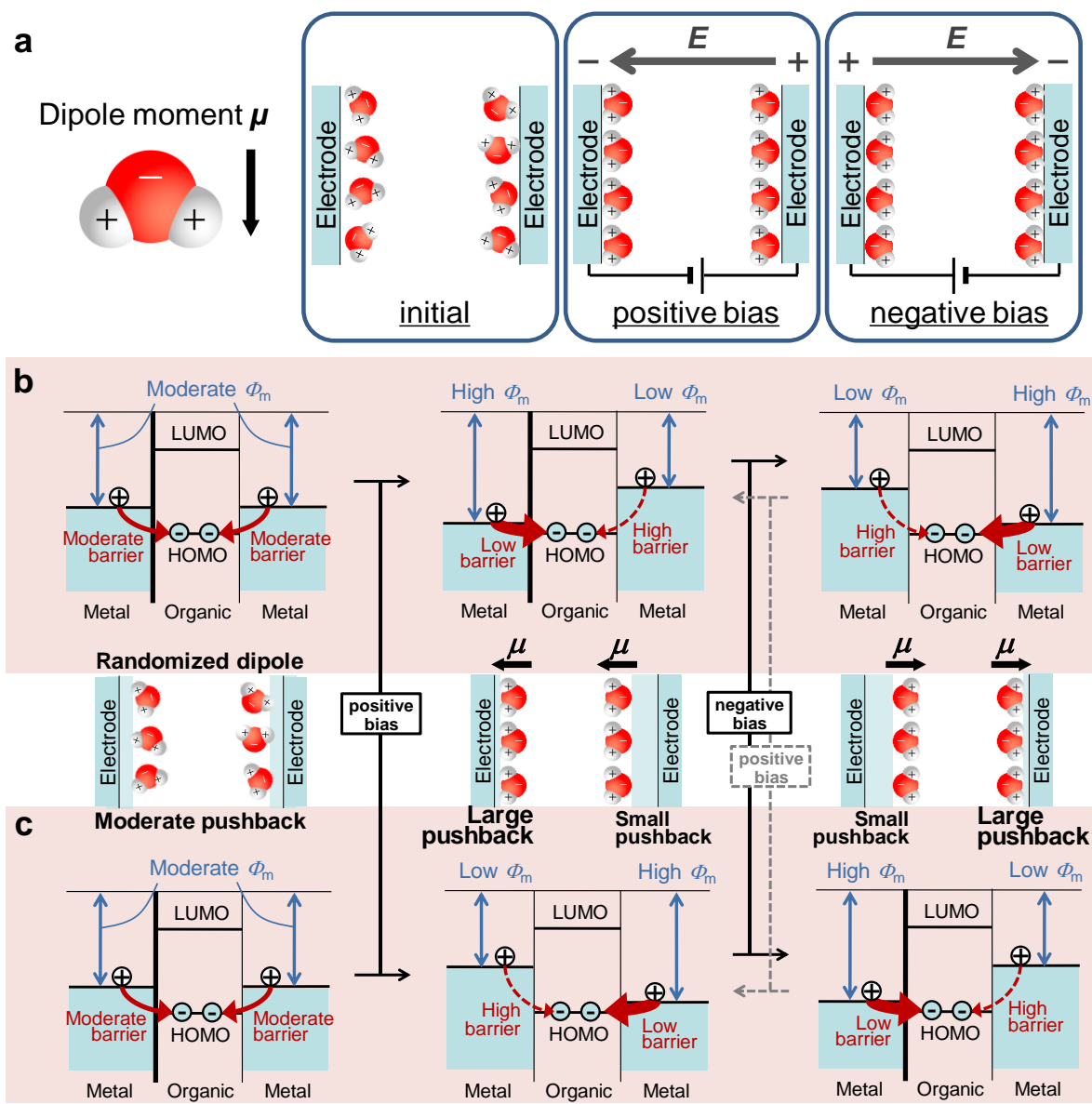

**Figure 5.** Mechanism of the dipolar switching. a) Orientational switching of $H_2O$ owing to the permanent electric dipole. The orientational change can be regulated by the direction of an external electric field. b) Energy diagrams expected for the dipole model, which are different from those deduced experimentally (Figure 2d). c) Energy diagrams expected for the push-back model, which are identical to those deduced experimentally (Figure 2d).

**A mechanism for operational instability of organic electronic devices** is proposed and experimentally confirmed. Polar molecules in air ($H_2O$) penetrate into the organic semiconductor layer and reach the electrode surface. The electric field generated by the voltage application to the electrode orients the electric dipole of the polar molecules, which effectively changes the charge-injection barrier height at the electrode/semiconductor interfaces.

R. Nouchi*

**Dipolar Switching of Charge-injection Barriers at Electrode/Semiconductor Interfaces as a Mechanism for Water-induced Instabilities of Organic Devices**

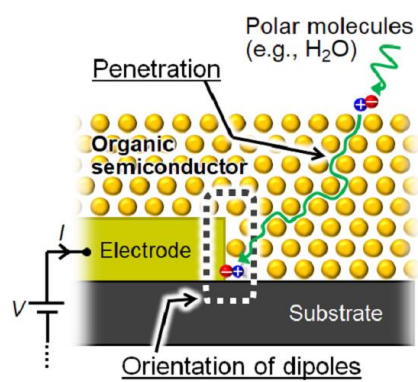